\documentstyle[times,epsfig,prl,aps,amssymb]{revtex}
\begin{document}
\draft
\twocolumn[\hsize\textwidth\columnwidth\hsize\csname
@twocolumnfalse\endcsname
\title{Global culture: A noise induced transition in finite systems}
\author{Konstantin Klemm$^{1,2,}$\cite{kk},
V\'{\i}ctor M. Egu\'{\i}luz$^{1,}$\cite{vme}, Ra\'ul Toral$^1$,
Maxi San Miguel$^1$}
\address{$^1$Instituto Mediterr\'aneo de Estudios Avanzados IMEDEA (CSIC-UIB),
E07071 Palma de Mallorca (Spain)\\
$^2$The Niels Bohr Institute, Blegdamsvej 17, DK2100 Copenhagen \O  (Denmark)}
\date{\today}
\maketitle

\begin{abstract}
We analyze the effect of cultural drift, modeled as noise, in Axelrod's
model for the dissemination of culture. The disordered multicultural
configurations are found to be metastable. This general result is proven rigorously 
in $d=1$, where the dynamics is described in terms of a Lyapunov
potential. In $d=2$, the dynamics is governed by the average relaxation
time $T$ of perturbations. Noise at a rate $r \lesssim T^{-1}$ induces
monocultural configurations, whereas $r \gtrsim T^{-1}$ sustains disorder. In
the thermodynamic limit, the relaxation time diverges and
global polarization persists in spite of a dynamics of local convergence.
\end{abstract}
]

Concepts and tools developed in the framework of Statistical and
Nonlinear Physics have been shown to be useful in identifying
general mechanisms behind collective behavior in social dynamics
\cite{Oliveira99}. Beyond the use of equilibrium ideas based on
minimization of a potential or optimization methods
\cite{Blume93}, models of interacting social agents, as for
example generalized voter models of opinion formation
\cite{Galam90}, fall within the context of the study of
nonequilibrium phase transitions in lattice models
\cite{Marro98}.
More generally, these are stochastic
spatial models also considered in population dynamics or
evolutionary biology \cite{Nowak92}.

Rather independently of this framework, but closely related in spirit,
is the model put forward by Axelrod to describe the dissemination of
culture among interacting agents in a society (or among societies)
\cite{Axelrod97}. Culture is here defined as a set of individual
attributes that are subject to social influence. The agents are placed
in the nodes of a square lattice, and the individual attributes of each
site are defined by a set of $F$ {\em features}, each taking one of $q$
possible {\em traits}. The dynamics of social influence takes into
account that ``the more similar an actor is to a neighbor, the more
likely the actor will adopt one of the neighbor's traits"
\cite{Axelrod97}. Specifically, two neighboring sites chosen at random
interact with a probability proportional to the number of common
features they already share. An interaction translates into switching
one of the different features of one site to the same trait of the
neighboring site. In principle, this dynamics tends to homogenize
neighbors, and hence to reach a completely ordered state (a {\sl
monocultural state}) in which each feature has the same trait in all
lattice sites. What Axelrod emphasizes about his model is that it
illustrates how the mechanism of local convergence can generate global
polarization: He finds that the system gets trapped in {\sl
multicultural states} with a number of different stable homogeneous
domains, called {\em cultural regions}. The number of domains in this
disordered state is used as a measure of cultural diversity. The
quantitative Statistical Mechanics analysis partially
modifies Axelrod's conclusion \cite{Castellano00}. Defining an order
parameter as the relative size of the largest homogeneous domain, one
finds a nonequilibrium order--disorder phase transition controlled by
the number of traits $q$, which measures the initial degree of disorder
of a random configuration. Symmetry breaking leading to dominance of a
given culture occurs for $q<q_c$, while the order parameter vanishes
for $q>q_c$. Therefore local convergence is efficient for $q<q_c$ and
global polarization only persists for $q>q_c$.

The analysis of \cite{Castellano00} corresponds to a zero-temperature
dynamics in which fluctuations are neglected. Here we address the issue
of the effects of random perturbations ({\sl noise}) following the
original suggestion of Axelrod, ``Perhaps the most interesting
extension and, at the same time, the most difficult one to analyze is
cultural drift [modeled as spontaneous change in a
trait]"\cite{Axelrod97,referee}. Axelrod also identified that a crucial
question in this case is to determine a characteristic time scale of
the dynamics. We find that this characteristic time scales with the
system size. A $q$-independent noise induced transition
\cite{transition} occurs when the noise rate $r$ is of the order of the
inverse of this time scale $T$: for $r \lesssim T^{-1}$ noise induces
ordered monocultural configuration because the disordered multicultural
states are metastable; for $r \gtrsim T^{-1}$ symmetry restoring by
noise is efficient and the disordered multicultural state prevails. In
the thermodynamic limit of large systems the condition $r \gtrsim
T^{-1} \rightarrow 0$ is satisfied for any finite $r$. In this limit,
the fundamental idea of local convergence generating global
polarization is recovered.

The model we study is defined \cite{Axelrod97} by considering $N$
agents as the sites of a lattice. The state of agent $i$ is a
vector of $F$ components (cultural features)
$(\sigma_{i1},\sigma_{i2},\cdots,\sigma_{iF})$. Each $\sigma_{if}$
is one of the $q$ integer values (cultural traits) $1,\dots,q$,
initially assigned independently and with equal probability $1/q$.
The time-discrete dynamics is defined as iterating the following
steps:
\begin{enumerate}
\item Select at random a pair of neighboring lattice sites $(i,j)$.
\item Calculate the {\em overlap} (number of shared features)
$l(i,j) = \sum_{f=1}^F \delta_{\sigma_{if},\sigma_{jf}}$.
\item If $0<l(i,j)<F$, the bond is said to be {\sl active} and sites
$i$ and $j$ interact with probability $l(i,j)/F$. In case of interaction,
choose $g$ randomly such that $\sigma_{ig}\neq\sigma_{jg}$ and set $\sigma_{ig}=\sigma_{jg}$.
\end{enumerate}

In any finite lattice the dynamics settles into an {\em absorbing}
state, characterized by the absence of active bonds. Obviousy all the
completely homogeneous states are absorbing. Inhomogeneous states
consisting of two or more homogeneous domains interconnected by bonds
with zero overlap are absorbing as well.

First we analyze the stability of these absorbing states. Before we
turn to the original two-dimensional geometry of the model, we consider
the one-dimensional lattice with nearest-neighbor interaction. For this
geometry the total negative overlap $V= -\sum_{i=1}^{N-1} l(i,i+1)$ is
a Lyapunov function: a function that never increases during the
dynamical process. Assume that at a given time step the interaction
across an active bond $(i,i+1)$ switches the state of one of the
features of site $i$. Then the overlap across that bond increases by
one unit. However, the overlap across the other bond $(i-1,i)$ of site
$i$ can at worst decrease by one unit. Taking into account that all
other terms in $V$ do not vary, we find that in any interaction, $V$
never increases. There is a multiplicity of $q^F$ ground states of $V$
that correspond to the homogeneous (ordered) configurations
$\sigma_{if}=\sigma_{jf}$ $\forall i,j$ and $f$, where the potential
$V$ takes its absolute minimum $V_{\rm min}=-N F$. Any other
(disordered) absorbing configuration is not the absolute minimum of $V$
and therefore is a metastable minimum of the Lyapunov potential.

\begin{figure}
\centerline{\epsfig{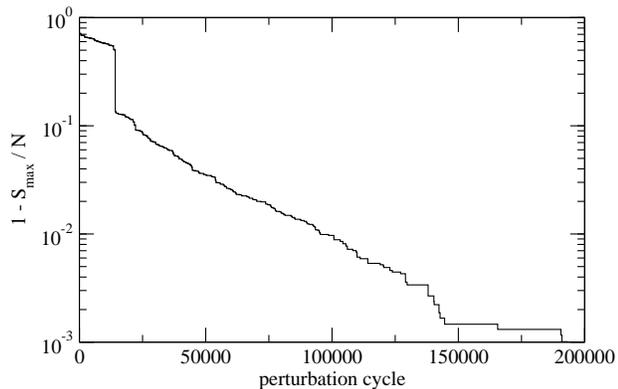}}
\caption{\label{fig1} Metastability of disordered absorbing
configurations: The order paramater $S_{\rm max}$ normalized with
systemsize $N$ is shown as a function of perturbation cycles for a
square lattice of $N=100^2$ sites using $F=10$ and $q=60$.}
\end{figure}

We have not found a parallel argument for $d$-dimensional lattices
($d>1$). However, by extensive numerical simulations we have verified
the metastability of disordered configurations for $d=2$ with nearest
neighbor interaction as well. The absorbing states are subject to {\em
single feature perturbations}, defined as randomly choosing
$i\in\{1,\dots,N\}$, $f\in\{1,\dots,F\}$ and $s\in\{1,\dots,q\}$ and
setting $\sigma_{if}=s$. Then the simulations are designed as follows:
(I) Draw a random initial configuration. (II) Run the dynamics by
iterating steps (1), (2) and (3), until an absorbing state is reached.
(III) Perform a single feature perturbation of the absorbing state and
resume at (II). We find that by these cycles of relaxation (step II)
and perturbation (step III) the system is driven to complete order,
where $\sigma_{if}=\sigma_{jf}$, $\forall i,j$ and $f$.  Thus, as in
the one-dimensional case, only the completely ordered configurations
are stable, all other absorbing configurations are merely metastable.
In Fig. \ref{fig1} we have plotted a typical evolution of the order
parameter considered in \cite{Castellano00}, that is, the size of the
largest homogeneous domain $S_{\rm max}$. We observe that, although
eventually its value could decrease, $S_{\rm max}$ typically increases
in each cycle of perturbation and relaxation \cite{noise2}.

We next consider the effect of cultural drift modeled as a random
perturbation acting continuously on the system (noise). This is
implemented by including a fourth step in the iterated loop of the
model:
\begin{enumerate}
\setcounter{enumi}{3}
\item With probability $r$, perform a single feature perturbation.
\end{enumerate}
Note that now the cultural imitation among sites and the perturbations are two processes
acting on the state vectors on time scales separated by the factor $r$. In contrast to the
previous scenario, the system is not
necessarily in an absorbing configuration when a perturbation occurs.

\begin{figure}
\centerline{\epsfig{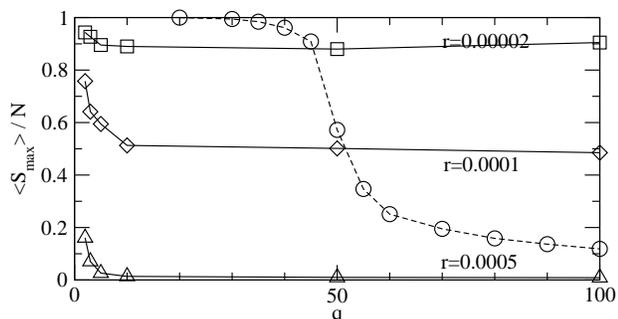}}
\caption{\label{fig2}  $\langle S_{\rm max}\rangle$ as a function of $q$ for
different values of $r$. The dashed line is for $r=0$ with a
transition at $q=q_c\approx 50$. The simulations used $N=50^2$ sites with $F=10$ features. Values $\langle S_{\rm max}\rangle$ in Figures \protect\ref{fig2}, \protect\ref{fig3} \protect\ref{fig4}, and \protect\ref{fig6} are averages
over 500 configurations. Measurements were taken after a relaxation time of $100N^2$ time steps.}
\end{figure}

\begin{figure}
\centerline{\epsfig{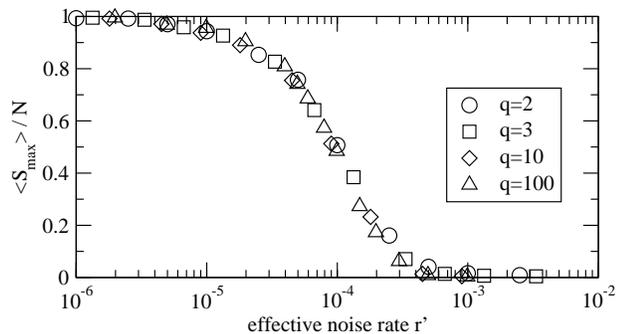}}
\caption{\label{fig3} $\langle S_{\rm max}\rangle$ as a function of the
effective noise rate $r'=r(1-1/q)$ for different values of $q$.
Simulations have been run in systems of size $N=50^2$ with
$F=10$.}
\end{figure}

Figure \ref{fig2} shows the average order parameter  $\langle S_{\rm
max}\rangle$ as a function of the number of traits $q$. In the case
$r=0$, we observe the expected transition from order to disorder as $q$
is increased \cite{Castellano00,remark1}. Strikingly, no such
transition is observed for noise rate $r \neq 0$. In fact, $\langle
S_{\rm max}\rangle$ is practically independent of $q$ except for small
values. This becomes clear by noticing that with probability $1/q$ a
perturbation does not change the configuration. Therefore considering the
effective noise rate $r^\prime = r(1-1/q)$ the data collapse onto a
single curve $\langle S_{\rm max}(r^\prime)\rangle$ (see
Fig.~\ref{fig3}). This curve identifies, for a fixed size of the
system, a continuous order--disorder transition controlled by the noise
rate. In addition, it shows that the transition has
universal scaling properties with respect to the value of $q$.

It is worth noting that the limit $r \to 0$ does not recover the
results for the original model without noise, corresponding to $r=0$.
The fact that for $r \to 0$ we find an ordered state for all values of
$q$ is linked to the metastable nature of the inhomogeneous absorbing
states previously discussed. Perturbations at a vanishingly small rate
are sufficient to allow the system to escape from these states.

The order--disorder transition is also observed in the distribution of
sizes of homogeneous domains. For small noise rates typically a single
cluster spans the whole system. As the noise rate is increased smaller
clusters become more and more abundant. At an intermediate value of the
noise rate the distribution of domain sizes follows a power law. The
exponent is approximately $-2$.

What causes the onset of disorder with increasing noise rate? For a
sufficiently small noise rate the system has enough time to relax to an
absorbing configuration between perturbations. Then the situation would
be comparable to the previously studied case of alternating
perturbation and relaxation: after a transient of reducing the disorder
from the initial condition like the one shown in Fig. 1, the system
will spend most of the time in one of the homogeneous configurations.
At variance with the case of alternating perturbation and
relaxation where the ordered state cannot be left once reached, there
will be, on long time scales, random jumps taking the system from a
homogeneous state to another homogeneous one. If the
noise rate is increased such that the typical time $1/r$ between
perturbations is shorter than the average relaxation time,
perturbations are ``accumulated'' in the system and disorder is built
up. According to this picture the average relaxation time $T$ of
perturbations of a homogeneous state sets the transition where $rT =
{\cal O} (1)$.

An approximate argument for the calculation of $T$ is as follows.
A single feature perturbation of an ordered state at time $t=0$
induces a ``damage'' of size $x(t=0)=1$ in one of the components.
In the following time steps the damage may spread until an ordered
state is reached again by $x(t)=0$ or $x(t)=N$. The probability
$D_x$ for $x$ to increase or decrease is the fraction of active
bonds, which again depends on $x$ but also on the current shape of
the damage cluster. This complication is avoided in a mean-field
description, where a bond exists between any two sites. Then the
number of active bonds is simply the number of pairs $(i,j)$ of
sites $i$ carrying the damage and sites $j$ not carrying the
damage. Given damage size $x$ we find $x(N-x)$ active out of
$N(N-1)/2$ bonds, yielding an update probability (for $N\gg
1$): $D_x = 2 N^{-2} x(N-x)$. The average first exit time, that is
the average number of time steps $\tau_{x}$ required to reach an
absorbing boundary ($x=0$ or $x=N$) when starting from $x(t=0)=x$,
fulfills the recursion relation\cite{GS}
\begin{equation} \label{taueq}
\tau_x = \frac{1}{2}D_x (\tau_{x-1} + \tau_{x+1}) + (1-D_x) \tau_x + 1~.
\end{equation}
with the boundary conditions $\tau_0=\tau_N=0$. This equation has the solution
\begin{equation}
\tau_x= -\frac{x}{N} \sum_{\xi=1}^{N-1} \frac{(N-\xi)}{D_\xi}
                    + \sum_{\xi=1}^{x-1} \frac{(x-\xi)}{D_\xi}~.
\end{equation}
The average relaxation time $T$ is the average first exit time for an
initial damage size $x=1$. Thus we obtain $
T = \tau_1 = N \left(\gamma + \ln(N-1) + {\cal O}(N^{-1})\right)$,
with Euler's constant $\gamma\approx0.577$. For large $N$ we approximate
\begin{equation}
T= N \ln(N),
\end{equation}
so that the average relaxation time of perturbations diverges with increasing system
size.

\begin{figure}
\centerline{\epsfig{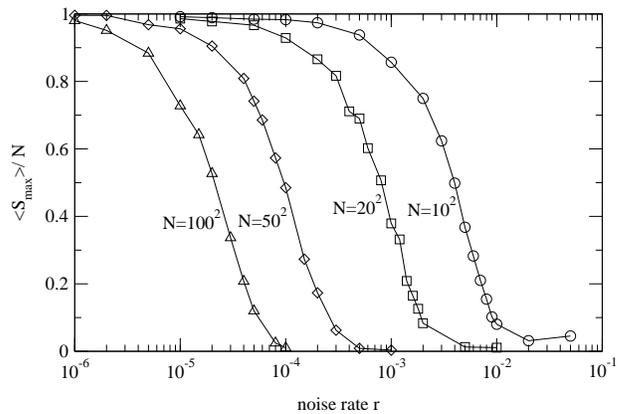}}
\caption{\label{fig4} $\langle S_{\rm max}\rangle$ as a function of noise rate
for different values of $N$. Simulations were performed with
$F=10$ and $q=100$. }
\end{figure}

\begin{figure}
\centerline{\epsfig{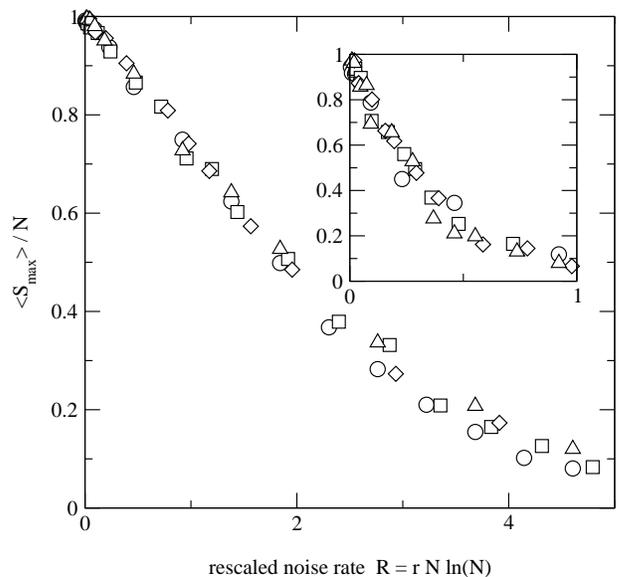}}
\caption{\label{fig6} $\langle S_{\rm max}\rangle$ plotted as a function of the
rescaled noise rate $R=rN\ln N$. Displayed symbols and model
parameters are the same as for Fig.~\protect\ref{fig4}. The inset
is the corresponding plot for the decoupled model, see the main
text. }
\end{figure}

This result, together with our argument that the transition occurs for
$rT = {\cal O} (1)$, suggests to consider systems of different size. We
always find a smooth transition from the ordered to the disordered
state as $r$ is increased, see Fig.~\ref{fig4}. The noise rate
necessary to induce disorder is seen to decrease with growing system
size. Fig.~\ref{fig6} confirms that $T$ is the relevant time scale for
the onset of disorder: The data of Fig.~\ref{fig4} collapse into a
single curve when plotted as a function of a rescaled noise rate
$R=rT=rN\ln(N)$ which incorporates noise rate $r$ and system size $N$.
Recalling our result for a fixed system size, we conclude that there is
a universal scaling form for the order parameter describing the
transition: $\langle S_{\rm max}(r,q,N)\rangle = \langle S_{\rm
max}(f(r,q,N)\rangle$ with $f(r,q,N) = r(1-1/q)N\ln(N)$.

We finally note that for the correct prediction of the characteristic
time scale $T$ and the change of behavior with noise rate $r$, the
dependence of the interaction probability on the sites' overlap need
not be taken into account. This motivates the introduction of a
``decoupled'' version of the dynamics in which a site always adopts the
trait of the chosen neighboring site independently of the number of
shared features. It turns out that the size of the largest cluster
follows the same dependence on the parameters $q$, $r>0$ and $N$ as in
the original model (inset of Fig~\ref{fig6}). This shows that in the
presence of noise our main results are insensitive to one of the basic
premises of Axelrod's model.

In summary, we have described the scaling properties of an
order--disorder transition induced by noise which occurs at a
size-dependent value of the noise rate $T^{-1}(N)$. For a finite system
and $r \lesssim T^{-1}$ ordered monocultural configurations are induced
by noise because disordered multicultural configurations are
metastable. In the thermodynamic limit $N\rightarrow\infty$, the
condition $r \gtrsim T^{-1}(N)\rightarrow 0$ is satisfied for
arbitrarily small noise rate and the system remains in a disordered or
multicultural state. In any case, cultural drift changes the nature of
the ordered and disordered states. In the ordered state the system is
not stuck in a single homogeneous configuration, but during long time
scales visits in succession a series of monocultural configurations.
Likewise, the disordered sate is not a frozen configuration but an
evolving state with noise sustained dynamics. Thus, the cultural drift
appears to be a relevant variable which drastically modifies the
dynamics of Axelrod's cultural model.

We acknowledge financial support from DGES (MCyT, Spain) under projects
BFM2000-1108 and BFM2001-0341-C02-01.


\begin{thebibliography}{99}

\bibitem[\dagger]{kk}e-mail: {\tt klemm@imedea.uib.es} 
\bibitem[\star]{vme}e-mail: {\tt victor@imedea.uib.es} 

\bibitem{Oliveira99}
S. M de Oliveira, P.M.C. de Oliveira, D. Stauffer , {\em
Nontraditional applications of computational Statistical Physics}
(B. G. Teubner Stuttgart, Leipzig, 1999).

\bibitem{Blume93}
L. Blume, Games and Econ. Behav. {\bf5}, 387 (1993); {\bf 11}, 111
(1995).

\bibitem{Galam90}
S. Galam, J. Stat. Phys. {\bf61}, 943 (1990); S. Galam, B.
Chopard, A. Masselot, M. Droz, Eur. Phys. J. B {\bf4}, 529 (1998).

\bibitem{Marro98}
J. Marro and R. Dickman, {\em Nonequilibrium phase transitions in
lattice models} (Cambridge University Press, Cambridge, 1998);
I. Dornic, H. Chat\'e, J. Chave, H. Hinrichsen, Phys. Rev. Lett.
{\bf87}, 045701 (2001).

\bibitem{Nowak92}
M. Nowak and R. May, Nature {\bf359}, 826 (1992); Int. Jour. of
Bif. and Chaos, {\bf3}, 35 (1993);
R. Durret, SIAM Review {\bf41}, 677 (1999).

\bibitem{Axelrod97}
R. Axelrod, J. Conflict Res. {\bf41}, 203 (1997). Reprinted in
\protect\cite{AxelrodBook}.

\bibitem{AxelrodBook}
R. Axelrod, {\em The Complexity of
Cooperation}, (Princeton University Press, Princeton, 1997).

\bibitem{Castellano00}
C. Castellano, M. Marsili, A Vespignani, Phys. Rev. Lett. {\bf85}, 3536 (2000)

\bibitem{referee} See also the discussion about the importance of
cultural drift in \cite{AxelrodBook}, pp. 145--147.

\bibitem{transition} We use the term {\sl transition} to refer to the
change of behavior observed in the system, although we will later show
that it is not a {\sl phase} transition in the sense of equilibrium 
Statistical Mechanics because it disappears in the thermodynamic
limit.

\bibitem{noise2} We note that metastability is here defined in terms of
a set of allowed perturbations: We have made similar numerical
simulations with perturbations where {\em all} features of a given site
are randomly assigned new traits. Then the system remains in the
disordered state only if $q$ exceeds a threshold value. At the
threshold the size of the largest cluster undergoes a transition from
zero to finite values.

\bibitem{remark1} The model analyzed in
Ref.~\protect\cite{Castellano00} is different from Axelrod's model as
in the initial condition the traits are not equally but Poisson
distributed. As we employ Axelrod's original model with uniform
distribution of traits, our result for $r=0$ confirms that the
order--disorder transition also occurs in the original model.

\bibitem{GS} G.R. Grimmett and D.R. Stirzaker, {\sl Probability and Random Processes}, Oxford Science Publications (1982).

\end{thebibliography}
\end{document}